\documentclass[aps,prd,twocolumn,showkeys,groupedaddress,nofootinbib]{revtex4-1}
\usepackage{graphicx}
\usepackage{subfig}
\usepackage{filecontents}

\begin{document}

\preprint{Proceedings of the $22^{nd}$ International Spin Symposium}

\title{Recursive Monte Carlo code for transversely polarized quark jet}

\author{ A. Kerbizi$^{\, 1}$, X. Artru$^{\, 2}$, Z. Belghobsi$^{\, 3}$, F. Bradamante$^{\, 1}$, A. Martin$^{\, 1}$ and E. Redouane Salah$^{\,4}$}

\affiliation{$^{1}$ {\small INFN Sezione di Trieste and Dipartimento di Fisica, Universit\`a di Trieste,}\\
{\small Via Valerio 2, 34127 Trieste, Italy}\\
$^{2}$ {\small Univ. Lyon, Universit\'e Lyon 1, CNRS/IN2P3,}
{\small Institut de Physique Nucl\'eaire de Lyon, 69622 Villeurbanne, France}
\\$^{3}$ {\small Laboratoire de Physique Th\'eorique, Facult\'e des Sciences Exactes et de l'Informatique,}\\ 
{\small Universit\'e Mohammed Seddik Ben Yahia,}\\
{\small B.P. 98 Ouled Aissa, 18000 Jijel, Algeria}\\
$^{4}$ {\small D\'epartement de Physique, Université\'e Mohamed Boudiaf, 28000 M'sila, Algeria}\\
}


\begin{abstract}
We propose a Monte Carlo code for the simulation of the fragmentation process of polarized quarks into pseudoscalar mesons. Such process is generated recursively once the flavour, the energy and the spin density matrix of the initial quark are specified, performing a cascade of splittings of the type $q\rightarrow h+q'$, where $q, q'$ indicate quarks and $h$ the hadron with flavour content $q\bar{q}'$. Each splitting is generated using a splitting distribution which has been calculated in a string fragmentation framework including, for the first time, the ${}^3P_0$ mechanism. This mechanism involves a "complex mass" parameter responsible for transverse spin effects, such as the Collins effect. Results for single hadron and hadron pair analyzing power are found to be in agreement with experimental results from SIDIS and $e^+e^-$ annihilation.
\end{abstract}


\maketitle

\section{Introduction}
Being a non perturbative process, the study of the fragmentation of quarks into hadrons requires the use of models. Among the variety of presently existing models \cite{models}, the particular class of recursive type has been extensively used in the simulation of high energy scattering processes where the decay of a system of color charges in singlet state needs to be handled. Of particular interest are the Feynman-Field model \cite{feynman-field} and the Lund Symmetric Model (LSM) \cite{lund}. The latter is the basis of the fragmentation part in the event generator Pythia \cite{pythia}, successful in the description of experimental data for different scattering processes not involving polarization.

In the string formalism, the chromoelectric field stretched between the two endpoint color charges, which are either a quark $q$ and an antiquark $\bar{q}$ in $e^+e^-$ annihilation or a quark $q$ and a diquark $qq$ in Deep Inelastic Scattering (DIS), is replaced by a classical string which evolves in spacetime. During its evolution the string is cut several times in different spacetime points due to quark-antiquark pairs tunneling out of the background field. After the pair production, quarks and antiquarks combine to form stable hadrons or resonances which subsequentely decay.
The quark spin degree of freedom in the fragmentation process, which is a non trivial complication as well as an essential ingredient, is not included in the existing generators of quark fragmentation processes. Having a complete generator is however needed for the preparation and the analysis of experiments where the hard scattering process involves polarized quarks. Among the effects originated by the quark spin the Collins effect \cite{collins} is of particular interest. It correlates the transverse polarization of the quark $q$ and the transverse momentum of the hadron $h$ in the process $q\uparrow \rightarrow h+X$. This correlation is described by the chiral odd function $H_{1q}^{\perp h}$ known as the Collins function. $H_{1q}^{\perp h}$ is interesting as a non perturbative object and it is used in quark polarimetry, being convoluted with the chiral-odd transversity PDF $h_1^q$ in Semi-Inclusive DIS (SIDIS).

The Monte Carlo code described here simulates the fragmentation of a polarized quark with flavour $u$, $d$ or $s$ and fixed energy. The quark jet is generated by recursive splittings $q\rightarrow h+q'$ where the four-momentum of the hadron $h$ is drawn according to the splitting distribution $F_{q',h,q}$ calculated in a string fragmentation framework, similar to the LSM, with quark spin developed by Artru and Belghobsi \cite{Dubna2013}. Here, the quark-antiquark pairs created in string cutting points are supposed to tunnel out of the vacuum in a ${}^3P_0$ state, i.e. in a triplet spin state $S=1$  with unit relative orbital angular momentum $L=1$, such that $J\equiv L+S=0$. It was already shown in a simplified quark multiperipheral model \cite{Dubna2009} that the ${}^3P_0$ wave function hypothesis predicts a Collins effect both for single hadron and for hadron pairs if a "complex mass" parameter is included. However, due to the complexity of the problem, in order to obtain quantitative predictions one has to rely on simulations. A different approach with respect to the string fragmentation model has been proposed by the authors of Ref. \cite{Matevosyan:2016fwi}.

The theoretical framework on which the code is based is described in Sections \ref{sec:code 1} and \ref{sec:code 2}. The results are summarized in Section \ref{sec:results}.

\section{The recursive algorithm}\label{sec:code 1}
The hadronization process $q_A+\bar{q}_B\rightarrow h_1+h_2+\dots + h_N$ is most naturally supposed to occur via a quark chain diagram and modelized as the set of splittings
\begin{eqnarray}
q_A\rightarrow h_1+q_2,\, q_2\rightarrow h_2 +q_3,\, \dots,\,q_N\rightarrow h_N +q_B
\end{eqnarray}
or equivalently as the recursive application of the elementary splitting
\begin{equation}
\label{eq: splitt elementary}
q\rightarrow h + q'
\end{equation}
where the flavour content of the hadron $h$ is $q\bar{q}'$. The production of vector mesons and baryons is not included in the present code.

In the following we indicate with $k$ and $k'$ the four-momenta of the quark $q$ and $q'$, with $p$ the four momentum of hadron $h$ and $p^{\pm}=p^0\pm p^z$ the lightcone momenta. We define also the longitudinal splitting variable $Z=p^+/k^+$, which is the forward lightcone momentum fraction of $q$ carried by the hadron $h$.

We assume the initiator quark $q_A$ to be traveling along the $\hat{\textbf{z}}$ axis, taken as the jet axis, with definite $k_A^+$ and no primordial transverse momentum. We take it completely polarized along the $\hat{\textbf{y}}$ axis.

Each hadron $h$ of mass $m_h$ is characterized by the four-momentum $p=(p^+,p^-,\textbf{p}_T)$. The lightcone momenta are linked by the mass-shell constraint $p^+p^-=\varepsilon^2_h$, where $\varepsilon^2_h=m_h^2+\textbf{p}_T^2$ is the hadrons transverse energy squared and $\textbf{p}_T$ is its transverse momentum with respect to the jet axis. In terms of the quark transverse momenta in the elementary splitting it is $\textbf{p}_T=\textbf{k}_T-\textbf{k}_T'$
due to momentum conservation.

The energy-momentum sharing between the hadron $h$ and the quark $q'$ in each splitting is given by the splitting distribution $F_{q',h,q}(Z,\textbf{p}_T,\textbf{k}_T)$ such that $F_{q',h,q}(Z,\textbf{p}_T,\textbf{k}_T)dZd^2\textbf{p}_T$ is the differential probability for $h$ to be emitted with lightcone momentum fraction $Z$ and transverse momentum $\textbf{p}_T$. We generate first $Z$ and then $\textbf{p}_T$. It makes it easier to take into account the dynamical correlations in the decay chain between the quarks transverse momenta $\textbf{k}_T$ and $\textbf{k}_T'$ as suggested in Ref. \cite{A.Z.E}.

Once the splitting distribution is calculated, the algorithm for the simulation of the jet generated by an initial polarized quark $q_A$ is the following
\begin{itemize}
\item[-] define the flavour, the energy-momentum and the spin density matrix $\rho(q)$ of $q_A\equiv q$
\begin{enumerate}
\item generate a new $q'\bar{q}'$ pair
\item form $h=q\bar{q}'$ and identify the meson type
\item generate $Z$ according to the $\textbf{p}_T$ integrated splitting distribution (see Section \ref{sec:code 2}) and calculate $p^+=Zk^+$
\item generate $\textbf{p}_T$ according to the splitting distribution at fixed $Z$
\item apply the mass shell constraint $p^+p^-=m_h^2+\textbf{p}_T^2$
\item test the  exit condition: if it is not satisfied continue to step $7$, otherwise the decay chain is ended
\item construct the hadrons four-momentum and store it
\item calculate $\textbf{k}_T'=\textbf{k}_T-\textbf{p}_T$ (for $q_A$, $\textbf{k}_T$ is known)
\item calculate the spin density matrix of quark $q'$.
\end{enumerate}
\item[-] repeat steps $1-9$ until the exit condition is satisfied
\end{itemize}

In step $1$ the generation of $s$ quarks is suppressed with respect to $u$ or $d$ quarks, by taking $P(u\bar{u}):P(d\bar{d}):P(s\bar{s})$ with probabilities $\alpha:\alpha:1-2\alpha$ such that $P(s\bar{s})/P(u\bar{u})=0.33$. This gives $\alpha\simeq 0.43$.

The meson identification at step $2$ uses the isospin wave function and also suppresses the $\eta^0$ meson production with repsect to $\pi^0$ to account for their mass difference. We have choosen $n(\eta^0)/n(\pi^0)\simeq 0.57$ as suggested in Ref. \cite{feynman-fieldw}.

As far as the momentum of the initiator quark is concerned, we focus on the DIS kinematics where the lightcone momentum of the hadronic system, composed by the virtual photon and the nucleon, in the nucleons rest frame is given by
\begin{equation}\label{eq: initial lightcone momenta}
P^{\pm}=M+\nu\pm\sqrt{\nu^2+Q^2}
\end{equation}
where $M$ is the nucleons mass. $P^{\pm}$ represent the reservoir of the lightcone momenta, $s=P^+P^-$ is the squared energy available for the fragmentation and $\nu=Q^2/2Mx_B$ is roughly the energy of the initial quark. Therefore $s$ is determined by the knowledge of the $x_B,\,Q^2$ values.

In our reference frame the quark $q_A$ travels along the forward lightcone and one can identify $k_A^+\equiv P^+$. We are interested in the fragmentation region of this initial quark. Therefore the remainder-jet initiated by $\bar{q}_B$, which in the DIS case is a diquark, can be neglected. The $\bar{q}_B$ particle travels along the backward lightcone with momentum $k_B^-\equiv P^-$. 

Let us suppose now $n$ hadrons have been generated. The remaining lightcone momenta are $P^{\pm}_{rem (n)}$ and after the generation of $h_n$ with momentum $(p_n^+,p_n^-,\textbf{p}_{T,n})$ it is, at step $n+1$,
\begin{eqnarray}
P^{+}_{rem (n+1)}=P^+_{rem (n)}-p^+_n \\ P^-_{rem (n+1)}=P^-_{rem (n)}-\frac{\varepsilon_{h_n}^2}{p^+_n} \\ \textbf{P}_{T,n+1}=\textbf{P}_{T,n}-\textbf{p}_{T,n+1}.
\end{eqnarray}
The remaining squared energy to be used in the generation of the next hadrons is $s_{n+1}=P_{rem(n+1)}^+P_{rem (n+1)}^--\textbf{P}_{T,n+1}^2$.
When the squared energy $s_{n+1}$ falls below a given mass $M_R^2$ the chain terminates (exit condition at step $6$). In our code it is $M_R=1.5\, GeV/c^2$ to leave enough energy for the production of a (not simulated) baryon. The observables investigated here are not sensitive to this value.

\section{The polarized splitting}\label{sec:code 2}
The main ingredient for the simulation of the fragmentation process is the splitting distribution which, in string fragmentation models, is almost fully determined by the principle of "left-right" symmetry (or quark chain reversal) \cite{lund}. This principle tells that the process should look the same if started from the $q_A$ side or from the $\bar{q}_B$ side.

If the quark polarization is taken into account, the splitting distribution becomes more complicated. It has been calculated \cite{Dubna2013} under the assumption that the $q\bar{q}$ pairs created in the string cutting points are produced in the ${}^3P_0$ state, obtaining
\begin{eqnarray}\label{eq: splitt spin}
\nonumber F_{q',h,q}(Z,\textbf{p}_T,\textbf{k}_T)dZd^2\textbf{p}_T\propto  \left(\frac{1-Z}{\varepsilon_h^2} \right)^a e^{-b_L\,\varepsilon_h^2/Z}\times \\Tr\left[g\lbrace q',h,q\rbrace\rho_{int}g^{\dagger}\lbrace q',h,q\rbrace\right] \frac{dZ}{Z}d^2\textbf{p}_T
\end{eqnarray}
where $a$ and $b_L$ are free parameters, playing the same role as $a$ and $b$ in the PYTHIA event generator. The former suppresses large $Z$ values while the latter is linked to an "intrinsic string fragility", i.e. the probability of having a string cutting point in the space-time area $dz\, dt$ of the string world sheet. $\rho_{int}$ is an intermediate spin matrix linked to the spin density matrix $\rho(q)$ of the quark $q$ through \cite{Dubna2013}
\begin{equation}
\rho_{int}=\frac{u^{-1/2}(\textbf{k}_T^2)\rho(q)u^{-1/2}(\textbf{k}_T^2)}{Tr[u^{-1/2}(\textbf{k}_T^2)\rho(q)u^{-1/2}(\textbf{k}_T^2)]}
\end{equation}
where $u(\textbf{k}_T^2)$ is a single-quark density in $\textbf{k}_T\otimes spin\,space$ (classical in $\textbf{k}_T$ but matrix density in spin). The expression for $u(\textbf{k}_T^2)$ as given in Ref. \cite{Dubna2013} is
\begin{eqnarray}\label{eq:u matrix}
\nonumber u(\textbf{k}_T^2)&=&\sum_{h,|s>}\int_0^1 \frac{dZ}{Z}\int d^2\,\textbf{p}_T \left(\frac{1-Z}{\varepsilon_h^2}\right)^a e^{-b_L\varepsilon_h^2/Z}\times\\
&\times& g^{\dagger}\lbrace q',h,q\rbrace g\lbrace q',h,q\rbrace
\end{eqnarray}
and the summation is on the final hadron type $h$ and its spin states $|s>$.

The quark transverse momenta and the spin matrices are included in the $g\lbrace q',h,q\rbrace$ matrix, which is  left-right symmetric, i.e. symmetric under $q\leftrightarrow q',\, h\leftrightarrow \bar{h}$.
For the spin matrix the ${}^3P_0$ mechanism suggests from \cite{Dubna2013}
\begin{eqnarray}\label{eq: g matrix}
g\lbrace q',h,q\rbrace &=& e^{-b_T(\textbf{k}_T^2+{\textbf{k}_T'}^2)/2} \varepsilon_h^{a}\times \gamma(\textbf{k}_T',\textbf{k}_T)
\end{eqnarray}
\begin{eqnarray}\label{eq:gamma matrix}
\nonumber \gamma(\textbf{k}_T',\textbf{k}_T)&=&(\mu_q'+\sigma_z \sigma\cdot\textbf{k}'_T)\sigma_z(\mu_q+\sigma_z\sigma\cdot\textbf{k}_T)\\
&=&\sigma_z\mu_q\mu_{q'}+\mu_{q'}\sigma\cdot \textbf{k}_T-\mu_q \sigma\cdot\textbf{k}_T' +\\
\nonumber &+&\sigma_z \left[\textbf{k}_T'\cdot\textbf{k}_T+i\sigma\cdot(\textbf{k}_T'\times\textbf{k}_T)\right]
\end{eqnarray}
where the terms in round brackets containing the complex masses $\mu_q$, $ \mu_{q'}$ are quark propagators, each one connecting two successive hadrons via a ${}^3P_0$ tunnel and the $\sigma_z$ matrix in between is the coupling for pseudoscalar meson emission. The parameter $b_T$ is linked to the width of the quark transverse momenta produced in the string cutting points.

The spin matrix $g\lbrace q',h,q\rbrace$ enters the splitting distribution in eq. (\ref{eq: splitt spin}) quadratically and in the limit of  small quark transverse momenta we have neglected the terms in $\textbf{k}_T\cdot\textbf{k}'_T$ and $\sigma\cdot(\textbf{k}_T\times\textbf{k}'_T)$ appearing in eq. (\ref{eq:gamma matrix}) which would give polynomial terms of degree four in eq. (\ref{eq: splitt spin}). Thus, we approximate eq. (\ref{eq:gamma matrix}) by
\begin{equation}\label{eq:gamma final}
\gamma(\textbf{k}_T',\textbf{k}_T)=\mu(\mu+\sigma_z\sigma\cdot \textbf{p}_T)
\end{equation}
Furthermore we have used only one complex mass parameter $\mu=(Re(\mu),Im(\mu))$ for all quark flavours.

Finally, the spin density matrix $\rho(qì)$ of the quark $q'$ (step $9$) is calculated according to \cite{Dubna2013}
\begin{equation}\label{eq: rho q'}
\rho(q')=\frac{g\lbrace q',h,q\rbrace \rho_{int}(q) g^{\dagger}\lbrace q',h,q\rbrace}{Tr\left[g\lbrace q',h,q\rbrace \rho_{int}(q) g^{\dagger}\lbrace q',h,q\rbrace\right]}
\end{equation}

\section{The simulation results}\label{sec:results}
As described above, the present simulation code handles the fragmentation of a polarized quark with fixed energy and without initial transverse momentum ("primordial" transverse momentum) into pseudoscalar mesons. It is a stand alone program not yet interfaced with existing event generators. As can be seen from eq. (\ref{eq: splitt spin}) and eq. (\ref{eq: g matrix}) there are $5$ free parameters in the model: $a$, $b_L$, $b_T$, $Re(\mu)$ and $Im(\mu)$. The values of $a$, $b_L$, $b_T$ and $|\mu|^2$ have been tuned comparing the simulation results for unpolarized fragmentation with experimental data on multiplicities of charged hadrons in SIDIS off unpolarized deuteron as function of $p_T^2$ \cite{makke2016transverse} and on unpolarized fragmentation functions \cite{kniehl2001testing}, until a satisfying qualitative agreement is found in a particular $[x_B,Q^2]$ kinematical region. The main slope of the multiplicities (in log scale) as function of $p_T^2$ is sensitive to $b_L$ and $b_T$ while the detailed shape for $p_T^2\rightarrow 0$ is sensitive to $|\mu|^2$. They are not affected by $a$, which changes the fragmentation functions at large fractional energies $z_h$ of the hadrons. Once the transverse polarization of the initiator quark is switched on, one can fix the ratio $Im(\mu)/Re(\mu)$ comparing the simulated and the experimental Collins asymmetries in $e^+e^-$. In this work we have used $a=0.9$, $b_L=0.5\,GeV^{-2}$, $b_T=5.17\,{(GeV/c)}^{-2}$, $Re(\mu)=0.42\, GeV/c^2$, $Im(\mu)=0.76\,GeV/c^2$ and compared the simulation results with experimental data for an initial $u$ quark.

\begin{figure}[tb]
  \includegraphics[width=0.8\linewidth]{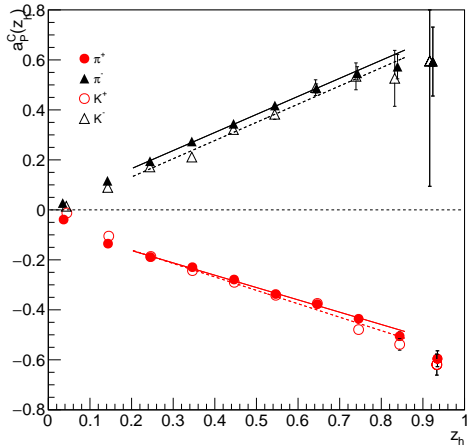}
 \caption{\small{Collins analyzing power as function of $z_h$ for charged mesons produced in simulated transversely polarized $u$ quark jets. The lines are linear fits to the simulation points. The cut $p_T>0.1\,(GeV/c)$ is applied for each meson.}}
  \label{fig:zh}
\end{figure}

Figure \ref{fig:zh} shows the Collins analyzing power $a_P^C$ for charged pions and kaons as function of $z_h=E_h/\nu$. $a_P^C$ is calculated as $2\langle \sin\phi_C\rangle$, where $\phi_C=\phi_h-\phi_{\textbf{s}_A}$ is the Collins angle, $\phi_h$ and $\phi_{\textbf{s}_A}$ are the azimuths of the hadron $h$ and of the transverse polarization of the initiator quark $q_A$. The mean values are given in Tab. \ref{tab:1h mean analyz power}. The main feature is that the analyzing power has opposite sign and almost equal magnitude for oppositely charged mesons, as qualitatively expected from the ${}^3P_0$ model and shown in the $e^+e^-$ data. Also, the analyzing power vanishes for small $z_h$ and is linear in the range $0.2<z_h<0.8$. The decrease of $a_P^C$ for small fractional energies is due to the fact that the memory of spin information is gradually lost when the rank increases. As $z_h$ becomes smaller the rank grows and the memory of the initial polarization decays \cite{Dubna2009}. As one can notice, the slope of the analyzing power for negative mesons, which are unfavoured in $u$ chains, is larger than the slope for positive ones. Furthermore the slope for $\pi^-$ and $K^-$ are similar. A linear dependence on $z_h$ is also suggested by BELLE data \cite{M.B.B} while it is not visible in the SIDIS data \cite{compassplb}, where the statistical uncertainties at large $z_h$ are large.

\begin{figure*}[tb]\centering
\begin{minipage}{.4\textwidth}
  \includegraphics[width=0.8\textwidth]{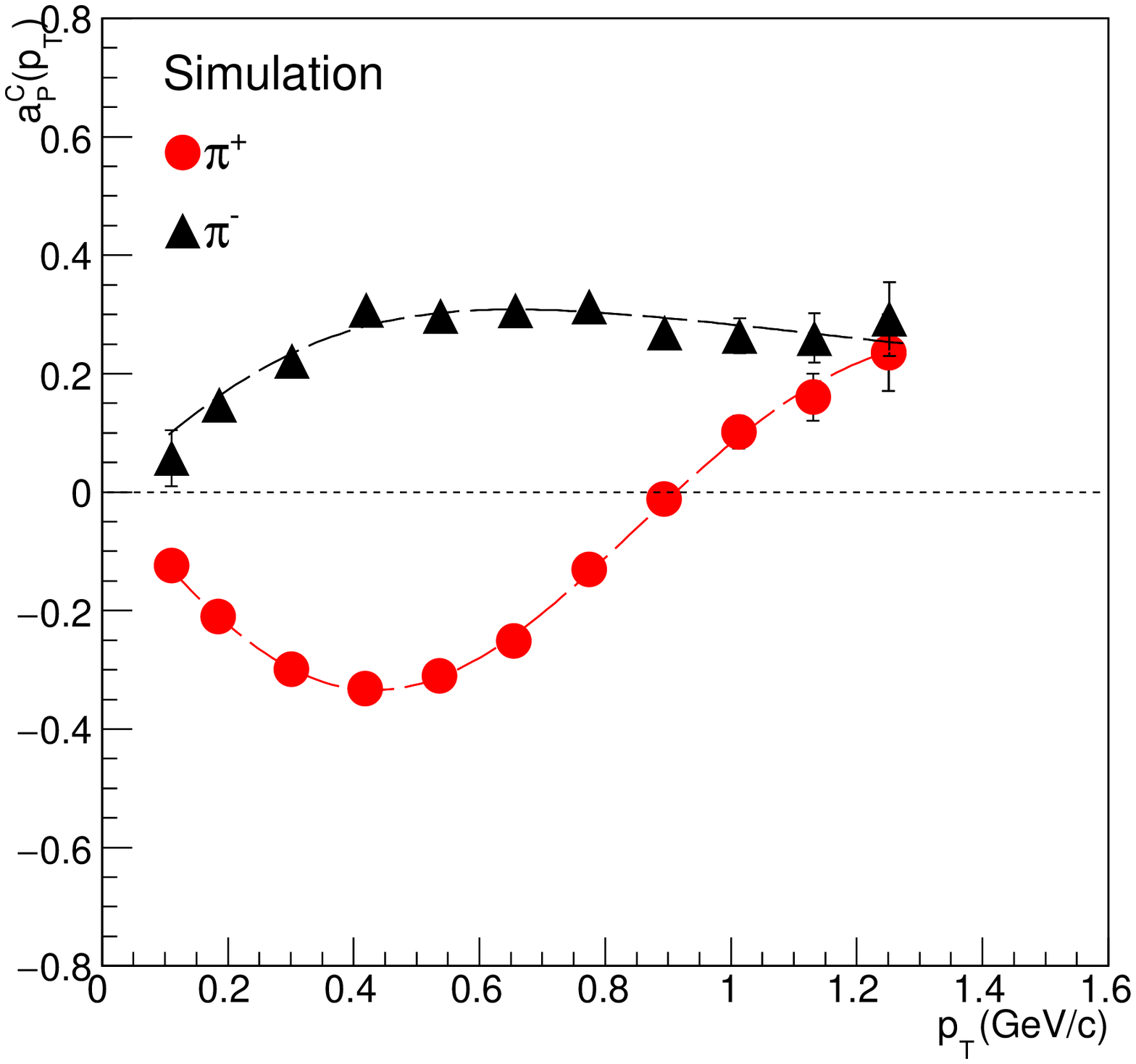}
\end{minipage}%
\begin{minipage}{.4\textwidth}
  \includegraphics[width=0.8\textwidth]{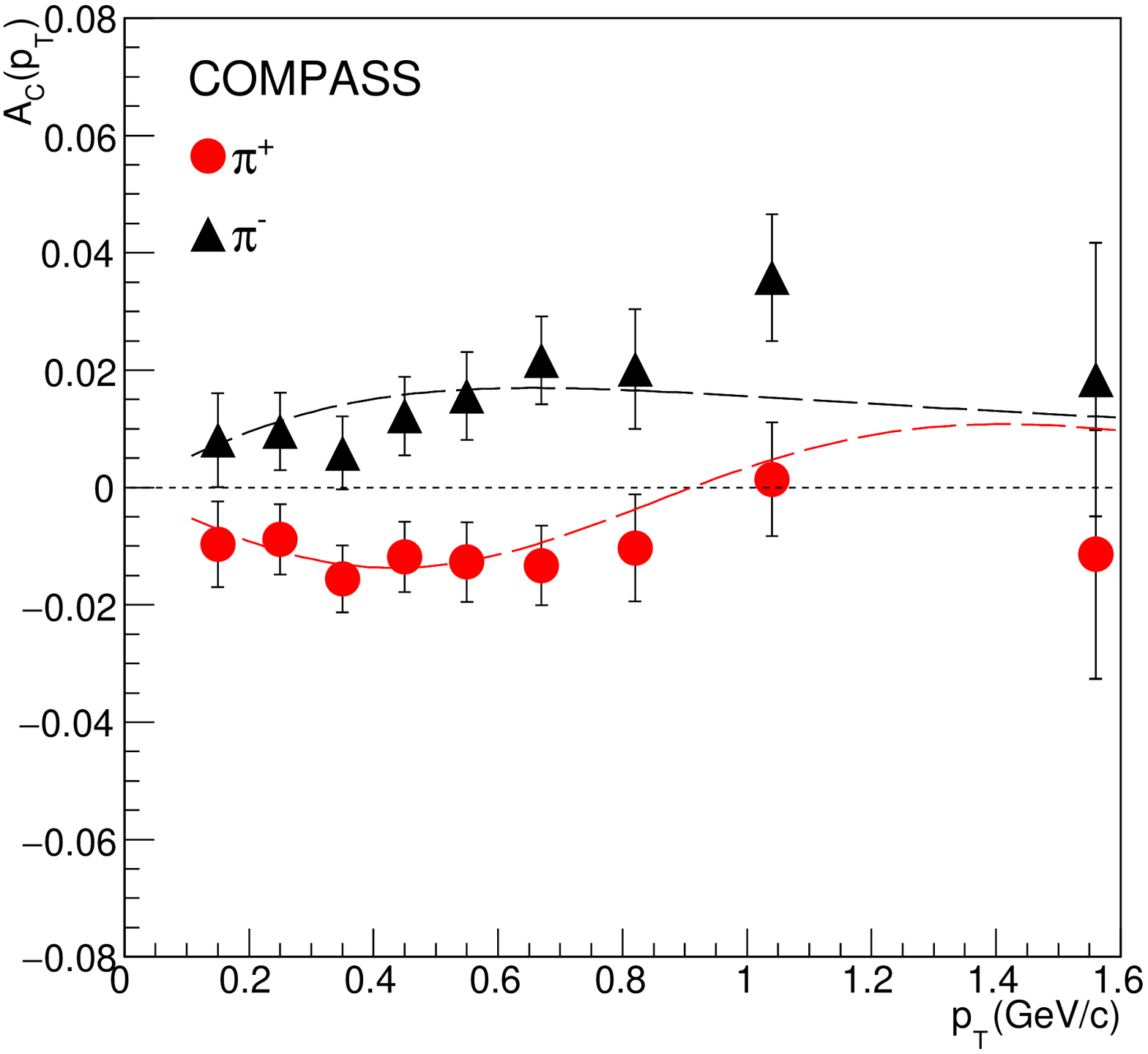}
\end{minipage}
\caption{\small{Left panel, Collins analyzing power as function of $p_T$ for charged pions produced in transversely polarized $u$ quark jets. The points are the result of the simulation. Right panel, Collins asymmetry measured in SIDIS off transversely polarized protons from Ref. \cite{compassplb}. The black and the red lines of the fits are obtained using eq. (\ref{eq:fit function 1}) and eq. (\ref{eq:fit function 2}) respectively.}}\label{fig:pt asymm}
\end{figure*}

The analyzing power as function of $p_T$ is shown in Fig. \ref{fig:pt asymm}. Clearly there are two different behaviours for positive and negative mesons. In the $\pi^-$ (and $K^-$) case the power law
\begin{equation}\label{eq:fit function 1}
a_P^C(p_T)_{h^-}=\frac{a_hp_T}{b_h^2+p_T^2}
\end{equation}
fits well the results, while the analyzing power for $\pi^+$ (and $K^+$) needs a stronger cutoff with $p_T$. In this case we have used the function
\begin{equation}\label{eq:fit function 2}
a_P^C(p_T)_{h^+}=\alpha_h p_T e^{-\beta_h p_T^2}(1-\gamma_h p_T^2)
\end{equation}
which decays faster for large transverse momenta. Note the change in sign of the analyzing power for positive pions for $p_T>0.9\,GeV/c$, as shown in Fig. \ref{fig:pt asymm}. It is due to the contribution of a second rank $\pi^+$ generated after a first rank $\pi^0$. Indeed, in the model, the second rank hadron has an asymmetry opposite to the first one and also a larger transverse momentum. The second term in eq. (\ref{eq:fit function 2}) is included to take into account such effect, which is expected to be washed out when the primordial transverse momentum is taken into account.

Similar shapes are observed in the Collins asymmetry for charged pions produced in SIDIS off transversely polarized protons measured by COMPASS \cite{compassplb} and shown in Fig. \ref{fig:pt asymm}, in the right panel. Here no change in sign for positive pions is observed but the trend of the asymmetry up to $p_T=0.9\,GeV/c$ is in fair agreement with simulations. Note that the different vertical scale values are due to the fact that the Collins asymmetry is the convolution of the $x$-integrated transversity PDF and $a_P^C$.

For jets initiated by transversely polarized $d$ quarks the analyzing powers for $\pi^+$ and $\pi^-$ from the simulation are identical but interchanged. This is not the case for $K^+$ and $K^-$, as can be easily understood in terms of their quark flavour content.

The same sample of simulated data has been used to investigate the properties of the analyzing power $a_P^{u\rightarrow \pi^+\pi^-}$ due to the Collins effect in the $\pi^+\pi^-$ pair production in the $u$ jet. Such analyzing power has been found to be qualitatively related to $a_P^C$ in a recent experimental work in SIDIS \cite{interplay} and its magnitude can be obtained by $e^+e^-$ data \cite{belle}.

In order to compare with the $e^+e^-$ data we have evaluated the quantity $\epsilon(M_{inv})\equiv\langle a_P^{u\rightarrow \pi^+\pi^-}\rangle a_P^{u\rightarrow \pi^+\pi^-}(M_{inv})$, where $M_{inv}$ is the $\pi^+\pi^-$ invariant mass and $\langle a_P^{u\rightarrow \pi^+\pi^-}\rangle$ is the analyzing power averaged over all the kinematical variables, including $M_{inv}$.

\begin{figure}[tb]\centering
\begin{minipage}{.4\textwidth}
  \includegraphics[width=0.8\textwidth]{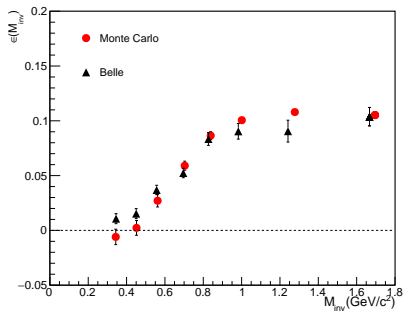}
\end{minipage} 
\caption{\small{Red circles represent the Monte Carlo calculation of $\epsilon(M_{inv})$ for pions produced in transversely polarized $u$ jets. Note that in the simulated data we ask for each pion of the pair $z_h>0.1$ and $p_T^2>0.1\, (GeV/c)^2$. The black triangles are $\epsilon(M_{inv})$ obtained from BELLE data \cite{belle}.}}\label{fig:minv pt2>0.1}
\end{figure}

Figure \ref{fig:minv pt2>0.1} shows $\epsilon(M_{inv})$ from the simulation (red circles) when $z_{h1,2}>0.1$ and $p_T^2>0.1\,(GeV/c)^2$. The black triangles represent the same quantity calculated from BELLE data \cite{belle}. The agreement between the two data sets is very striking.
In the simulation the analyzing power $a_P^{C,u\rightarrow \pi^+\pi^-}$ has been estimated using $2\langle\sin(\phi_B-\phi_{\textbf{s}_A})\rangle$ where $\phi_B$ is the azimuth of the vector $\textbf{p}_{1T}-\textbf{p}_{2T}$, $\textbf{p}_{1}$ ($\textbf{p}_{2}$) being the momentum of the $\pi^+(\pi^-)$ of the pair. 
Both in the simulation and in the data the analyzing power shows a saturation for large values of the invariant mass while for small values it decreases.

It has to be noted that the Monte Carlo data sample has a different invariant mass spectrum with respect to BELLE data. From the error bars one may notice that in the BELLE data sample the statistics is larger around the $\rho$ meson peak while in simulations the invariant mass spectrum is shifted towards large values. We remind that in the simulation program there are no resonances.

\begin{figure*}[tb]\centering
\begin{minipage}{.4\textwidth}
  \includegraphics[width=0.8\textwidth]{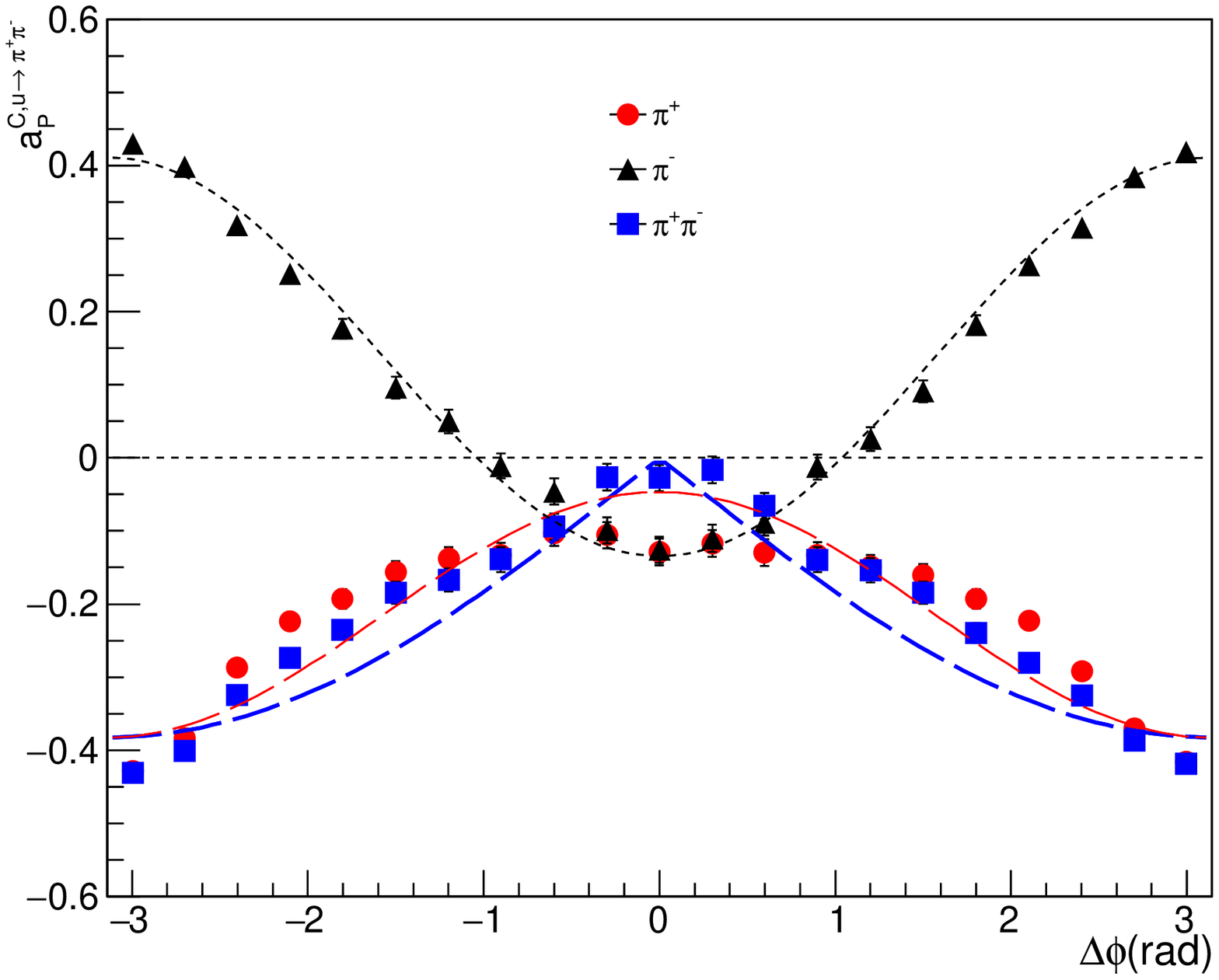}
\end{minipage}%
\begin{minipage}{.4\textwidth}
  \includegraphics[width=0.8\textwidth]{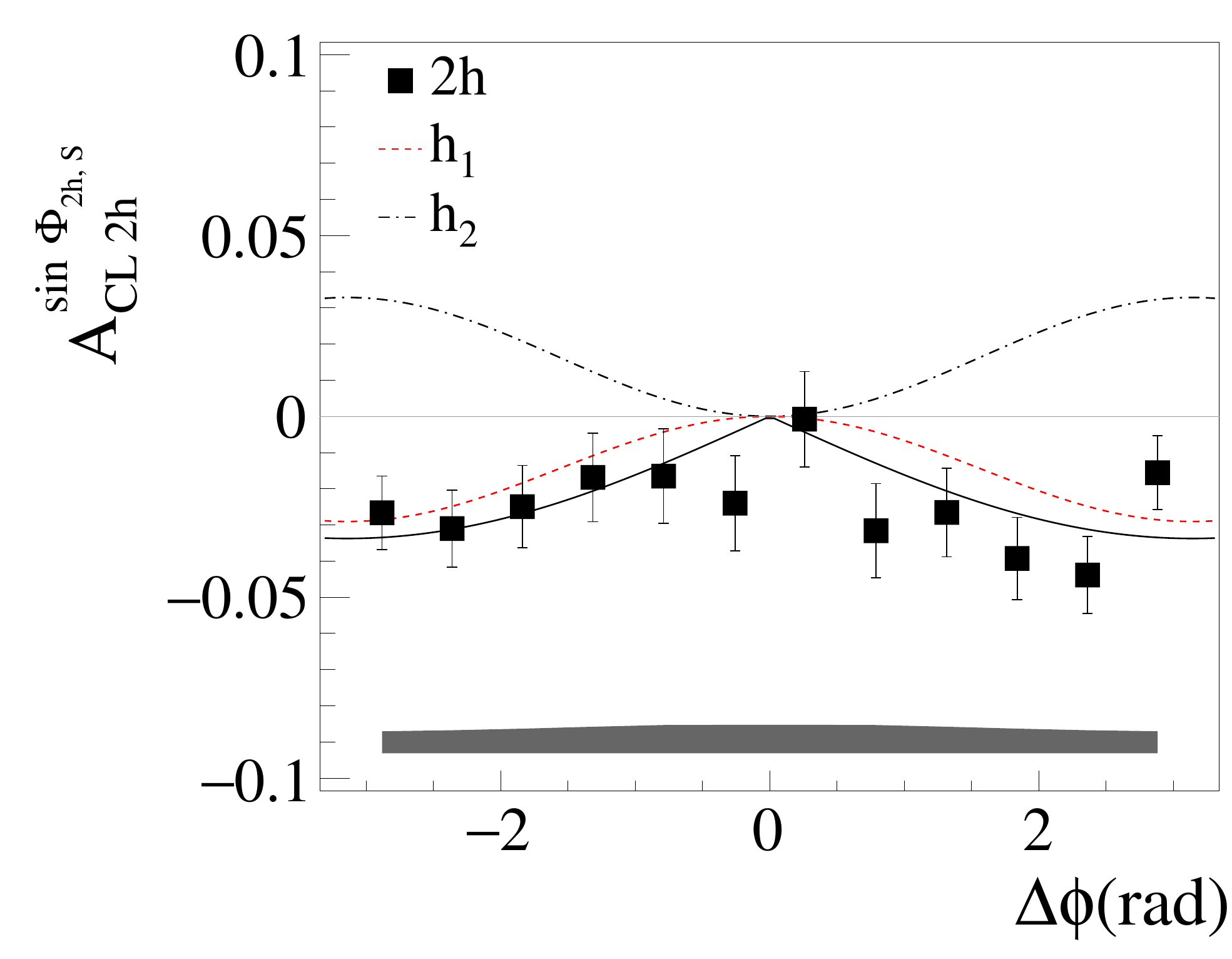}
\end{minipage}
\caption{\small{On the left panel (in blue) is shown the analyzing power for the Collins effect of $\pi^+\pi^-$ pairs produced in the same transversely polarized $u$ jet as function of the relative azimuthal angle $\Delta\phi$, as defined in Ref. \cite{interplay}. The red circles and the black triangles are the "Collins Like" asymmetries , as defined in Ref. \cite{interplay}, for charged pions and the dashed black and red curves are the corresponding fits. On the right panel is shown the same plot from COMPASS data as presented in Ref. \cite{interplay}.}}\label{fig:delta phi}
\end{figure*}

Finally we have studied the relationship between the Collins and the di-hadron analyzing powers for $\pi^+\pi^-$ pairs in the same $u$ quark jet, as function of the relative azimuthal angle $\Delta\phi=\phi_1-\phi_2$ where the index $1$ ($2$) refers to $\pi^+$ ($\pi^-$), following what was done by the COMPASS Collaboration \cite{interplay}. Here we define $a_P^{u\rightarrow \pi^+\pi^-}$ as the mean value $2\langle \sin(\phi_{2h}-\phi_{\textbf{s}_A})\rangle$, where $\phi_{2h}$ is the azimuth of the vector $\hat{\textbf{p}}_{1T}-\hat{\textbf{p}}_{2T}$ and $\hat{\textbf{p}}_T\equiv \textbf{p}_T/|\textbf{p}_T|$. 
The blue points in Fig.\ref{fig:delta phi} (left) represent the di-hadron analyzing power $a_P^{C,u\rightarrow \pi^+\pi^-}$ calculated in the Monte Carlo as function of $\Delta\phi=\phi_1-\phi_2$. The blue curve is the result of the fit with the function $c\sqrt{2(1-\cos\Delta\phi)}$ as suggested in Ref. \cite{interplay} and it describes the trend of the Monte Carlo points.
The plot at the right in Fig. \ref{fig:delta phi} shows the asymmetry $A_{CL,2h}^{\sin\phi_{2h,S}}$ from Ref. \cite{interplay} as measured in COMPASS. As can be seen, the agreement is good and the main features are the same in real data and simulation. Note that the $A_{CL,2h}^{\sin\phi_{2h,S}}$ asymmetry is a factor of $0.1$ less than $a_P^{C,u\rightarrow \pi^+\pi^-}$ due to the transversity integral.
The slight up-down disymmetry for $\pi^+$ and $\pi^-$ in the simulated data is due to the different values of the analyzing power for $\pi^+$ and $\pi^-$, as shown in Tab. \ref{tab:1h mean analyz power}.

\begin{table}[b]
\caption{\label{tab:1h mean analyz power}\small{Mean value of the analyzing powers shown in Fig.\ref{fig:zh} for positive and negative charges. The cuts $z_h>0.2$ and $p_T>0.1$ have been applied.}}
\begin{ruledtabular}
\begin{tabular}{l*{6}{c}r}
$\langle a_P^C\rangle$         & $h^+$ & $h^-$   \\
\hline
$\pi$& $-0.254\pm 0.002$ & $0.270\pm 0.003$   \\
$K$ & $-0.268\pm 0.004$ & $0.230\pm 0.005$\\
\end{tabular}
\end{ruledtabular}
\end{table}

\section{Conclusions}
We have developed a stand alone Monte Carlo code for the simulation of the fragmentation process of a transversely polarized quark ($u$, $d$ or $s$) of fixed energy. The theoretical framework is provided by the string fragmentation model where the quark-antiquark pairs in the string cutting points are produced according to the ${}^3P_0$ mechanism. The quark spin is included through spin density matrices and propagated along the decay chain according to the ${}^3P_0$ rules.

With respect to the Lund Symmetric Model, this model requires an additional complex mass parameter whose imaginary part directly affects the single hadron Collins asymmetry. The three free parameters present in the string fragmentation framework and the absolute value of the complex mass have been tuned by comparison with unpolarized experimental SIDIS data.
The results of the simulation show a Collins effect of opposite sign for oppositely charged mesons. The dependence on the kinematical variables has been investigated, finding a good qualitative agreement with experimental data. A clear Collins effect for hadron pairs of opposite sign in the same jet is also obtained from the same simulated data. This effect is compared to BELLE and COMPASS asymmetries finding again a satisfactory agreement.

The results are very promising and the work is continuing. In particular in the future the code will be updated including resonances and their decays. An interface to presently existing event generators is also planned in order to make more quantitative comparisons with the existing data.

\bibliography{kerbizi_bibliography}

\end{document}